\documentclass[12pt]{article}
\usepackage{graphicx}
\usepackage{amsmath}
\begin{document}

\title{Q-balls constructed of spinors in Lagrangians with $SU(2)$
symmetry}

\author{Athanasios Prikas}

\date{}

\maketitle

Physics Department, National Technical University, Zografou
Campus, 157 80 Athens, Greece.\footnote{e-mail:
aprikas@central.ntua.gr}

\begin{abstract} In the present work we investigate the existence and
stability properties of q-balls which consist of a couple of
scalar fields, forming an $SU(2)$ doublet in a Lagrangian with a
global $SU(2)$ symmetry. We find that these spinors can form a
localized and stable field configuration, if they rotate in their
internal $SU(2)$ space. We find the energy and charge of the
soliton in both thin and thick-wall approximation and we prove
its stability against decaying to free particles. We also find
the asymptotic forms of the scalar and gauge field and the energy
and charge of the configuration when the $SU(2)$ symmetry is
local. The only assumption is the smallness of the coupling
constant $g$. Using numerical methods we prove the stability of
the q-ball in the local case.
\end{abstract}

PACS number(s): 11.27.+d, 11.10.Lm

\newpage

\section{Introduction}

Q-balls are non-topological solitons appearing in Lagrangians of
scalar fields with a global $U(1)$ or $SO(2)$ symmetry. Though
they appeared as a rather mathematical object \cite{math},
through a series of papers they revealed their physical
properties \cite{Col1,Col2,Col3}. Q-balls can be observed at, or
near, the minimum of the $\sqrt{U/{(|\Phi|)}^2}$ quantity, where
$U$ is the potential and $\Phi$ is the scalar field. At the above
minimum one can prove that the energy is minimized with respect
to the charge. The minimum of this quantity should be less than
the particle mass, in order the soliton to be stable. Much more
interest was concentrated on the subject when their possible
existence in supersymmetric extensions of the Standard Model was
considered \cite{susy1,susy2,susy3,susy4}, where the $U(1)$ charge
is for example the baryon number of the supersymmetric partners of
baryons. Flat $U(1)$ directions in the potential of such
extensions can offer a plausible explanation to the problem of
baryogenesis (\cite{bar1,bar2}).

Another type of q-ball type solutions has been studied, namely the
so-called non-abelian q-balls \cite{nonab1,nonab2} where we have
scalar fields in a Lagrangian with a global $SU(3)$ or $SO(3)$
symmetry usually and the fields belong to the adjoint
representation of the symmetry group. Such particles may
correspond to the sgauginos in $N=2$ supersymmetric theories.
Also, q-balls with a local $U(1)$ symmetry have been studied in
\cite{gauged1}.

In a recent series of papers \cite{flatsu21,flatsu22,flatsu23}, a
great number of $SU(2)$ flat directions in the superpotential of
the Minimal Supersymmetric Standard Model and a rich phenomenology
concerning them have been investigated. So the plausible question
is to ask if a new type of q-balls can appear in such directions,
enriching the particle phenomenology. In the present work we try
to do the following three things: a)We present the general
formalism concerning q-balls when the fields transform according
to the doublet representation of the symmetry group because this
case has not been studied and may turn out to be useful in the
search of q-solitons composed of supersymmetric partners of the
usual leptons. b)We present the formalism of gauged q-balls which
describe scalar particles with ``weak" interactions. c)We fully
investigate the properties of the above solitons both analytically
and numerically in the so-called thin- and thick-wall
approximation. Our main interest is to verify their stability when
varying the gauge coupling, the frequency with which the soliton
field rotates within its internal $SU(2)$ space, or other
phenomenological parameters.

\section{Global $\mathbf{SU(2)}$ symmetry.\\ Thin-wall approximation}

We regard a Lagrangian
\begin{equation}\label{1}
L={\partial}_{\mu}{\vec{\varphi}}^{\dag}{\partial}^{\mu}
\vec{\varphi}-U({\vec{\varphi}}^{\dag}\vec{\varphi})\ ,
\end{equation}
characterized by a global $SU(2)$ symmetry concerning the couple
of scalar fields:
\begin{equation}\label{2}
\vec{\varphi}=\left(\begin{array}{c}
  {\varphi}_1 \\
  {\varphi}_2
\end{array}\right)\ .
\end{equation}
The energy is provided by the equation
\begin{equation}\label{3}
\textrm{E}=\int{d^3x(\nabla \vec{\varphi}^{\dag}\nabla
\vec{\varphi}+ \dot{\vec{\varphi}}^{\dag}\dot{\vec{\varphi}}+U)}\
.
\end{equation}
The currents are three as the generators of the algebra
considered.  These currents are
\begin{eqnarray}\label{4}
j^0_{\alpha}=(\begin{array}{cc}
  \dot{\vec{\varphi}}^{\dag} &
  \dot{\vec{\varphi}}^T
\end{array}) \left(\begin{array}{cc}
  \frac{\imath}{2}{\tau}_{\alpha} & 0 \\
  0 & -\frac{\imath}{2}{\tau}_{\alpha}^{\ast}
\end{array}\right)\left(\begin{array}{c}
  \vec{\varphi} \\
   {\vec{\varphi}}^{\ast}
\end{array}\right)\ ,
\end{eqnarray}
where ${\tau}_{\alpha}$ are the Pauli matrices:
\begin{equation}\label{5}
{\tau}_1=\left(\begin{array}{cc}
  0 & 1 \\
  1 & 0
\end{array}\right) \hspace{2em} {\tau}_2=\left(\begin{array}{cc}
  0 & -\imath \\
  \imath & 0
\end{array}\right)  \hspace{2em} {\tau}_3=\left(\begin{array}{cc}
  1 & 0 \\
  0 & -1
\end{array}\right)\ .
\end{equation}

We choose
\begin{equation}\label{6}
\vec{\varphi}(\vec{r}
,t)=\exp\left(\frac{\imath}{2}{\tau}_{\alpha}{\varepsilon}_{\alpha}t\right)
\vec{\sigma}(\vec{r})\ .
\end{equation}
In the so-called thin-wall approximation we choose
$\vec{\sigma}(\vec{r})$ a step-function, with a constant value
within a volume $V$ and zero outside. We also choose a
spherically symmetric configuration, i.e.
$\vec{\sigma}(\vec{r})=\vec{\sigma}(r)$ so as to minimize the
contribution of the surface effects to the total energy. So the
energy can be written
\begin{equation}\label{7}
\textrm{E}=\dot{\vec{\varphi}}^{\dag}\dot{\vec{\varphi}}V+UV
\end{equation}
and using the chosen field configuration:
\begin{equation}\label{8}
\textrm{E}=\frac{1}{4}({\varepsilon}_1^2+{\varepsilon}_2^2+{\varepsilon}_3^2)
{\vec{\sigma}}^{\dag}\vec{\sigma}V+UV\ .
\end{equation}

We can now define the charges in the above thin-wall
approximation where we take
\begin{equation}\label{9}
Q_i=\frac{1}{2}{\varepsilon}_i
{\vec{\sigma}}^{\dag}\vec{\sigma}V\ .
\end{equation}
where the definitions:
\begin{equation}\label{10}
Q_{\alpha}=\int d^3xj^0_{\alpha}
\end{equation}
and
\begin{equation}\label{11}
\mathbf{Q} \equiv (Q_1,Q_2,Q_3)
\end{equation}
hold. We now want to write energy as a function of the charges. It
is a matter of simple algebra to show that:
\begin{equation}\label{12}
\textrm{E}=\frac{{\mathbf{Q}}^2}{{\vec{\sigma}}^{\dagger}\vec{\sigma}V}+UV\
.
\end{equation}
Regarding energy as a function of the volume and minimizing it
with respect to the volume, we take that the minimum is at:
\begin{equation}\label{13}
V=\sqrt{\frac{\mathbf{Q}^2}{{\vec{\sigma}}^{\dagger}\vec{\sigma}U}}
\end{equation}
and the energy takes the form:
\begin{equation}\label{14}
\textrm{E}=\sqrt{{\mathbf{Q}^2}}\sqrt{\frac{4U}{{\vec{\sigma}}^{\dagger}\vec{\sigma}}}\
.
\end{equation}

The energy of the free particles carrying charge
$\sqrt{{\mathbf{Q}^2}}$ is $2m\cdot\sqrt{{\mathbf{Q}^2}}$. (It is
a matter of simple algebra to prove this. The physical
interpretation is obvious because we now have two particles, each
with mass $m$.) So if the quantity
$(U/{\vec{\sigma}}^{\dag}\vec{\sigma})^{1/2}$ is less than the
mass of the free particles then the energy of the (\ref{14})
field configuration is less than the energy of the free waves and
the soliton is preserved stable against decaying into free
particles. We know that $\sqrt{{\mathbf{Q}}^2}$, quantity is both
conserved, as every single current is conserved, and $SU(2)$
invariant. We now investigate a potential of fourth power in
$|\varphi|\equiv {({\vec{\varphi}}^{\dag}\vec{\varphi})}^{1/2}$,
namely:
\begin{equation}\label{15}
U=m^2{|\varphi|}^2-
\frac{2\alpha}{3}{|\varphi|}^3+\frac{\beta}{2}{|\varphi|}^4\ .
\end{equation}
This is the more general potential of fourth order in the fields.
The theory is non-renormalizable but this is not a serious problem
as we can regard the Lagrangian as a phenomenological one. This
potential can be proved very useful in a supersymmetric theory.
Making the rescallings $|\varphi|\rightarrow
(m^2/\alpha)|\varphi|$, $x\rightarrow x/m$ and defining $B\equiv
\beta m^2/\alpha$ we take
\begin{equation}\label{16}
U={|\varphi|}^2-\frac{2}{3}{|\varphi|}^3+\frac{B}{2}{|\varphi|}^4\
.
\end{equation}
Minimizing the ${(4U/{|\varphi|}^2)}^{1/2}$ quantity with respect
to the field and calling it
$\sqrt{{\boldsymbol{\varepsilon}}_0^2}$ we can take that the
minimum is at
\begin{equation}\label{101}
|\varphi|=2/3B\ .
\end{equation}
Then:
\begin{equation}\label{102}
\sqrt{{\boldsymbol{\varepsilon}}_0^2}=2\sqrt{1-2/9B}\ ,
\end{equation}
\begin{equation}\label{103}
\mathbf{Q}=(1/2){(2/3B)}^2V\boldsymbol{\varepsilon}\ ,
\end{equation}
\begin{equation}\label{104}
\textrm{E}={(2/3B)}^2V\sqrt{{\boldsymbol{\varepsilon}}_0^2}\sqrt{1-2/9B}\
,
\end{equation}
where $\boldsymbol{\varepsilon}\equiv
({\varepsilon}_1,{\varepsilon}_2,{\varepsilon}_3)$. It is easy
then to verify that $\textrm{E}=
\sqrt{{\boldsymbol{\varepsilon}}_0^2}\sqrt{{\mathbf{Q}}^2}$.

\section{Local $\mathbf{SU(2)}$ case. General setting}

The Lagrangian for the case with a local $SU(2)$ symmetry is
written
\begin{equation}\label{25}
L={\vec{\varphi}}^{\dag}\left({\overleftarrow{\partial}}^{\mu}
+\imath g\frac{\boldsymbol{\tau}\cdot{\mathbf{A}}^{\mu}}{2}\right)
\left({\partial}_{\mu} -\imath
g\frac{\boldsymbol{\tau}\cdot{\mathbf{A}}_{\mu}}{2}\right)\vec{\varphi}
-U({\vec{\varphi}}^{\dag}
\vec{\varphi})-\frac{1}{4}{\mathbf{F}}^{\mu\nu}{\mathbf{F}}_{\mu\nu}\
,
\end{equation}
where:
\begin{equation}\label{26}
{\mathbf{F}}^{\mu\nu}={\partial}^{\mu}{\mathbf{A}}^{\nu}-
{\partial}^{\nu}{\mathbf{A}}^{\mu}+g{\mathbf{A}}^{\mu}\times{\mathbf{A}}^{\nu}\
.
\end{equation}
and $g$ is the coupling constant.

We now regard: a)Spherically symmetric fields in order to minimize
the contribution to the total energy of the so-called surface
effects. b)Static gauge fields. c)Only the zero component of the
gauge field differing from zero, the other components being equal
to zero. This is not a gauge-fixing but a statement that the
gauge field configuration is static and, consequently, the
magnetic fields absent. d)${\mathbf{A}}_0$ finite everywhere and
zero at infinity. If we also use the initial ansatz for the scalar
field we take for the Lagrangian:
\begin{equation}\label{27}
L=\frac{1}{4}{\vec{\sigma}}^{\dag}{(\boldsymbol{\varepsilon}
-g{\mathbf{A}}_0)}^2\vec{\sigma}-U({\vec{\sigma}}^{\dag}
\vec{\sigma})+\frac{1}{2}{({\partial}_a{\mathbf{A}}_0)}^2+
({\partial}_a{\vec{\sigma}}^{\dag})({\partial}^a\vec{\sigma})\ ,
\end{equation}
where Latin indices run from 1 to 3 and Greek indices run from 0
to 3.

Defining
\begin{equation}\label{28}
\boldsymbol{\theta}=\boldsymbol{\varepsilon}-g{\mathbf{A}}_0\ ,
\end{equation}
regarding $\boldsymbol{\varepsilon}$ as a constant and after a
little algebra we take
\begin{equation}\label{29}
\textrm{L}=4\pi\int r^2dr\left[\frac{1}{4}{\vec{\sigma}}^{\dag}
\vec{\sigma}{\boldsymbol{\theta}}^2- U({\vec{\sigma}}^{\dag}
\vec{\sigma})+\frac{1}{2g^2}{\boldsymbol{{\theta}'}}^2
-{{\vec{\sigma}}^{\dag'}}\vec{\sigma}' \right]\ ,
\end{equation}
where the prime denotes the derivative with respect to the radius
$r$ and the last term can be absent if $\vec{\sigma}(r)$ is a
step-function. The Euler-Lagrange equations for the above fields
$\vec{\sigma}$, $\boldsymbol{\theta}$, are
\begin{equation}\label{30}
\vec{\sigma}''+\frac{2}{r}\vec{\sigma}'+
\frac{1}{4}{\boldsymbol{\theta}}^2\vec{\sigma}- \frac{\partial
U}{\partial {\vec{\sigma}}^{\dag}}=0\ ,
\end{equation}
\begin{equation}\label{31}
\boldsymbol{\theta}''+\frac{2}{r}\boldsymbol{\theta}'-\frac{1}{2}g^2({\vec{\sigma}}^{\dag}
\vec{\sigma})\boldsymbol{\theta}=0\ .
\end{equation}
The total energy of the field configuration can be written
\begin{equation}\label{32}
\textrm{E}=4\pi\int r^2dr\left[\frac{1}{4}{\vec{\sigma}}^{\dag}
\vec{\sigma}{\boldsymbol{\theta}}^2+ U({\vec{\sigma}}^{\dag}
\vec{\sigma})+\frac{1}{2g^2}{\boldsymbol{{\theta}'}}^2
+{\vec{\sigma}}^{\dag'}{\vec{\sigma}}' \right]\ .
\end{equation}

Now we will find the total charge. A component of the current is
defined as:
\begin{equation}\label{33}
{j_{\alpha}}_0=\left[\begin{array}{cc}
  \frac{\partial L}{\partial({\partial}_0\vec{\varphi})} &
\frac{\partial L}{\partial({\partial}_0{\vec{\varphi}}^{\ast})}
\end{array}\right] \left[\begin{array}{cc}
  \frac{\imath}{2}{\tau}_{\alpha} & 0 \\
  0 & -\frac{\imath}{2}{\tau}_{\alpha}^{\ast}
\end{array}\right]\left[\begin{array}{c}
  \vec{\varphi} \\
   {\vec{\varphi}}^{\ast}
\end{array}\right]\ ,
\end{equation}
which in the special case of the local symmetry can be written
\begin{equation}\label{34}
{j_{\alpha}}_0=\left[\begin{array}{cc}
 {\vec{\varphi}}^{\dag}\left({\overleftarrow{\partial}}_0
+\imath g\frac{\boldsymbol{\tau}\cdot{\mathbf{A}}_0}{2}\right),  &
{\vec{\varphi}}^T\left({\partial}_0 -\imath
g\frac{{\boldsymbol{\tau}}^{\ast}\cdot{\mathbf{A}}_0}{2}\right)
\end{array}\right] \left[\begin{array}{cc}
  \frac{\imath}{2}{\tau}_{\alpha} & 0 \\
  0 & -\frac{\imath}{2}{\tau}_{\alpha}^{\ast}
\end{array}\right]\left[\begin{array}{c}
  \vec{\varphi} \\
   {\vec{\varphi}}^{\ast}
\end{array}\right]\ .
\end{equation}
After some algebra we take
\begin{equation}\label{35}
Q_i=2\pi \int r^2dr{\theta}_i{\vec{\sigma}}^{\dag}\vec{\sigma}\ .
\end{equation}

Another useful relation can be obtained from equation \ref{31}:
\begin{equation}\label{36}
(r^2\boldsymbol{\theta}')'=\frac{1}{2}g^2({\vec{\sigma}}^{\dag}
\vec{\sigma})\boldsymbol{\theta}r^2\ .
\end{equation}
By the definition of $\boldsymbol{\theta}$ and by substituting
eq. \ref{35} in \ref{34} we can take an asymptotic relation
connecting ${\theta}_i$, or equivalently ${A_i}_0$ and $Q_i$, when
$r\rightarrow \infty$. The relation is:
\begin{equation}\label{37}
{\theta}_i={\varepsilon}_i-g^2\frac{Q_i}{4\pi r}\ .
\end{equation}
Also for large $r$, $\vec{\sigma}$ is small, so as to have a
localized solution of the equations of motion, so
$U({\vec{\sigma}}^{\dag} \vec{\sigma})\cong
m^2{\vec{\sigma}}^{\dag} \vec{\sigma}$ and equation (\ref{30})
takes the form:
\begin{equation}\label{38}
\vec{\sigma}''+\frac{2}{r}\vec{\sigma}'+
\vec{\sigma}\left(\frac{1}{4}{\boldsymbol{\varepsilon}}^2-m^2\right)=0\
,
\end{equation}
the solution of which has the asymptotic form:
${\vec{\sigma}}_i\propto
\exp\left(-r\sqrt{m^2-\frac{1}{4}{\boldsymbol{\varepsilon}}^2}\right)/r$,
with $\vec{\sigma}_i$ the $i$ component of the $\vec{\sigma}$
doublet. In order to avoid oscillatory solutions we demand that
$\frac{1}{4}{\boldsymbol{\varepsilon}}^2<m^2$. Another useful
relation can be taken from the asymptotic forms of $\vec{\sigma}$
and $\boldsymbol{\theta}$ and from relations (\ref{32}) and
(\ref{35}):$$
\textrm{E}=\frac{1}{2}\boldsymbol{\varepsilon}\cdot\mathbf{Q}+4\pi\int
r^2dr[U({\vec{\sigma}}^{\dag}\vec{\sigma})+
{\vec{\sigma}^{\dag'}}\vec{\sigma}']\ ,$$ where $\mathbf{Q}\equiv
(Q_1,Q_2,Q_3)$ as defined above. So, if this energy is less than
the energy of the free particles the soliton is stable against
decaying to free particles.

We will now try to find a solution corresponding to a quite large
soliton in the thin-wall approximation. We will also take some
general properties of the soliton. The more convenient method for
solving the Euler-Lagrange equations is to regard the ``matter"
field $\sigma$ as a constant within a certain volume. In order to
minimize the surface contribution to the total energy we regard
the field configuration as spherical. The final scope of our work
is to find an expression of the energy with respect to the charge,
as we do in the global $SU(2)$ case. We will minimize the energy
firstly with respect to the radius of the soliton and then with
respect to the field value, as is the usual practice in the
treatment of q-balls. We will use the following approximations:
a)The thin-wall approximation for the matter field $\vec{\sigma}$.
b)The assumption of small coupling constant, $g$. We will not use
the approximate asymptotic relation for the energy given above as
in \cite{gauged1} where the case of gauged $U(1)$ q-balls was
treated, but the exact one, (equation \ref{32}). Let the matter
have a spatially constant value ${\sigma}_0$ within a sphere of
radius $R$. We will use eq. (31) so as to find ${\theta}_i(r)$. It
is easy to find that:
\begin{equation}\label{39}
{\theta}_i(r)=\frac{\left({\varepsilon}_i-\frac{g^2Q_i}{4\pi R
}\right)R\sinh(g\Sigma r)}{r\sinh(g\Sigma R)}\ , \hspace{1em}
r\leq R
\end{equation}
and:
\begin{equation}\label{40}
{\theta}_i(r)={\varepsilon}_i-\frac{g^2Q_i}{4\pi r}\ ,
\hspace{1em} r\geq R\ ,
\end{equation}
where $\Sigma \equiv {[{(1/2)
\vec{\sigma}}^{\dag}\vec{\sigma}]}^{1/2}$. It is easy to see that
the gauge field at large distances of the soliton origin has the
form of a field generated from a spherically symmetric localized
distribution carrying $SU(2)$ charge, without magnetic fields.
Eq. \ref{35} gives
\begin{equation}\label{41}
{\varepsilon}_i=\frac{g^3Q_i\Sigma}{4\pi} \left(\frac{1}{g\Sigma
R-\tanh(g\Sigma R)}\right)\ .
\end{equation}
Inserting ${\varepsilon}_i$ from the above equation in the
relations \ref{40}-\ref{41} and substituting the result in the
exact expression of the energy we find:
\begin{eqnarray}\label{42}
E=\frac{g^2{\mathbf{Q}}^2}{8\pi R}+
\{3[-1+\exp(2gR\Sigma)]g^2{\mathbf{Q}}^2- \nonumber\\
32[-1+\exp(2gR\Sigma)]
{\pi}^2R^4U+32[1+\exp(2gR\Sigma)]g{\pi}^2R^5\Sigma U\} \nonumber\\
{\{24\pi R[1+gR\Sigma+\exp(2gR\Sigma)(-1+gR\Sigma)]\}}^{-1}\ .
\end{eqnarray}
In the above expression the first term is the energy contribution
of the gauge field outside the q-ball. The other terms give the
energy of the interior of the soliton. We differentiate the above
expression with respect to the radius of the configuration and we
set the result equal to zero. In order to solve the transcendental
equation that we take, we expand it in a power series of the
coupling constant $g$. The equation that we have to solve is:
\begin{equation}\label{43}
-\frac{9{\mathbf{Q}}^2}{8\pi R^4{\Sigma}^2}+4\pi R^2U-
\frac{3{\mathbf{Q}}^2 g^2}{20\pi R^2}=0\ .
\end{equation}
The solution to the equation should also be expanded in a power
series of $g$:
\begin{equation}\label{44}
R=\frac{1}{2}{\left(\frac{3}{\pi}\right)}^{1/3}\frac{{(2{\mathbf{Q}}^2
{\Sigma}^4U^2)}^{1/6}}{\Sigma\sqrt{U}}+ \frac{{\mathbf{Q}}^2
{\Sigma}^4Ug^2}{\Sigma\sqrt{U}60\pi
{(2{\mathbf{Q}}^2{\Sigma}^4U^2)}^{1/2}}\ .
\end{equation}
Inserting the above expression into the expression of energy and
expanding again the energy with respect to $g$ we find
\begin{equation}\label{45}
\textrm{E}={\left({\mathbf{Q}}^2\frac{2U}{{\Sigma}^2}\right)}^{1/2}\left[1+
\frac{g^2{({\mathbf{Q}}^2)}^{1/3}
{\left(\frac{1}{4\pi}\frac{3{\Sigma}^2}{\sqrt{2U}}\right)}^{2/3}}{5}
\right]\ .
\end{equation}

The limit of the above expression when $g\rightarrow 0$ is what
we expect from the global case. The final step is to minimize the
energy with respect to the field $\Sigma$. Let ${\Sigma}_0$ be
the value of the field minimizing the energy in the global case.
Then the value that minimizes the energy in the local case is
slightly different from the first one, as one can expect when the
coupling constant is small and is given by:
\begin{equation}\label{46}
\Sigma={\Sigma}_0\left(1-\frac{g^2}{5}
{\left(\frac{2{\mathbf{Q}}^2}{3{\pi}^2}
\frac{{({\boldsymbol{{\varepsilon}_0}}^2)}^2}{{\Sigma}^4_0}
\right)}^{1/3}{\left[{\left[\frac{{\partial}^2}{\partial
{\Sigma}^2}\left(\frac{U}{{\Sigma}^2}\right)\right]}_{\Sigma={\Sigma}_0}\right]}^{-1}\right)\
,
\end{equation}
where:
\begin{equation}\label{47}
{\boldsymbol{\varepsilon}}_0^2
\equiv\frac{2U({\Sigma}_0)}{{\Sigma}^2_0}\ ,
\end{equation}
the value of the ``frequency" that minimizes the energy in the
$g\rightarrow 0$ case. Let
$$k\equiv\left[{\left[\frac{{\partial}^2}{\partial
{\Sigma}^2}\left(\frac{U}{{\Sigma}^2}\right)\right]}_{\Sigma={\Sigma}_0}\right]\
.
$$ If we substitute eq. \ref{46} into \ref{45} and keep only
the $g^2$ terms of a Taylor expansion we take
\begin{equation}\label{48}
\textrm{E}=\sqrt{{\mathbf{Q}}^2\frac{2U}{{\Sigma}_0^2}}\left[1+\frac{
3{\Sigma}_0^{4/3}\kappa{({\mathbf{Q}}^2)}^{1/3}+4{\left[\frac{
{({\boldsymbol{\varepsilon}}^2_0)}^2{\mathbf{Q}}^2}{{\Sigma}^4_0}\right]}^{1/3}}
{10\ast{12}^{1/3}\kappa{\pi}^{2/3}U^{1/3}}g^2\right]\ .
\end{equation}
In conclusion, we minimized the energy with respect to the radius
of the field configuration (eq. \ref{45}) and the field
$\vec{\sigma}$ (eq. \ref{48}). The energy seems to be larger than
in the global case due to the terms depending on $g^2$. There is
actually a slight decrease in the energy with the coupling
constant. This result will be verified numerically. But the
energy also depends on the charge. This total charge,
$\sqrt{{\mathbf{Q}}^2}$, may be different in the local case from
the global one. Intuitively, we expect that this charge (particle
number) is smaller in the case of the local symmetry due to the
electrostatic-type repulsion. This result will also be verified by
numerical analysis.

\section{Numerical results}

The Euler-Lagrange equations of the matter and the gauge field are
eqs. \ref{30}, \ref{31}. The energy and the soliton charge are
given from the equations \ref{32} and \ref{35} respectively. In
the global case we put $g=0$ and
$\boldsymbol{\theta}=\boldsymbol{\varepsilon}$. The soliton charge
is half the energy of the free particles with the same charge if
the mass is unity as we have seen.

\begin{figure}
\centering
\includegraphics{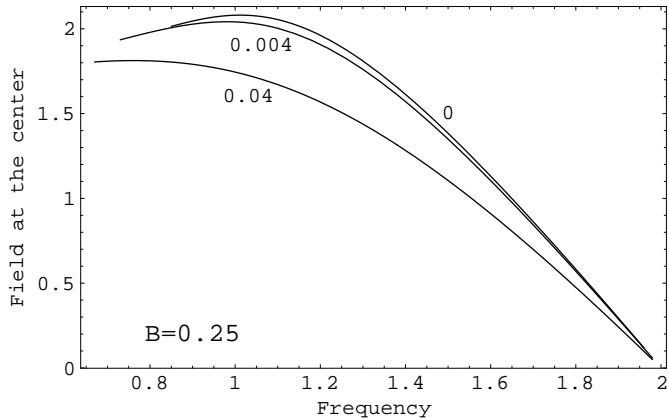}
\caption{The central value of the field as a function of
$\sqrt{{\boldsymbol{\varepsilon}}^2}$ for three different values
of the coupling constant with $B=0.25$.} \label{figure1}
\end{figure}

\begin{figure}
\centering
\includegraphics{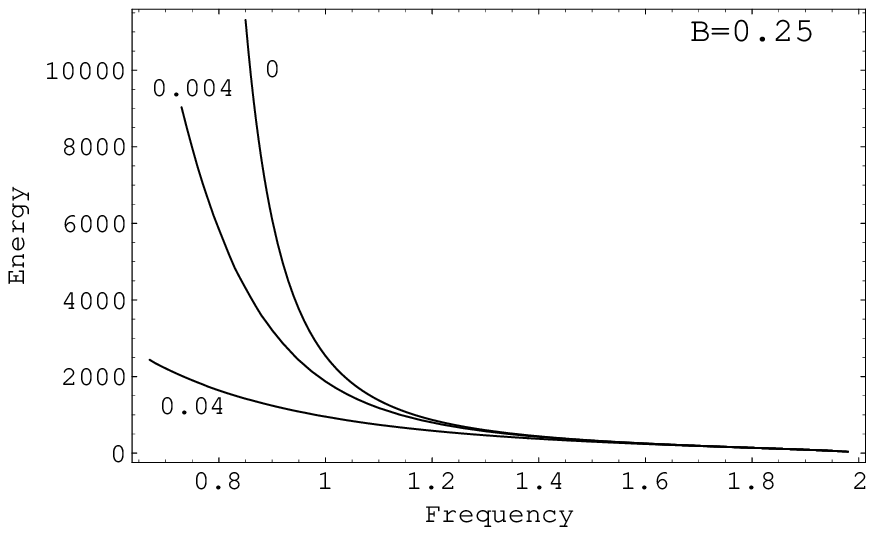}
\caption{The energy of the soliton as a function of
$\sqrt{{\boldsymbol{\varepsilon}}^2}$ for three different values
of the coupling constant with $B=0.25$.} \label{figure2}
\end{figure}

\begin{figure}
\centering
\includegraphics{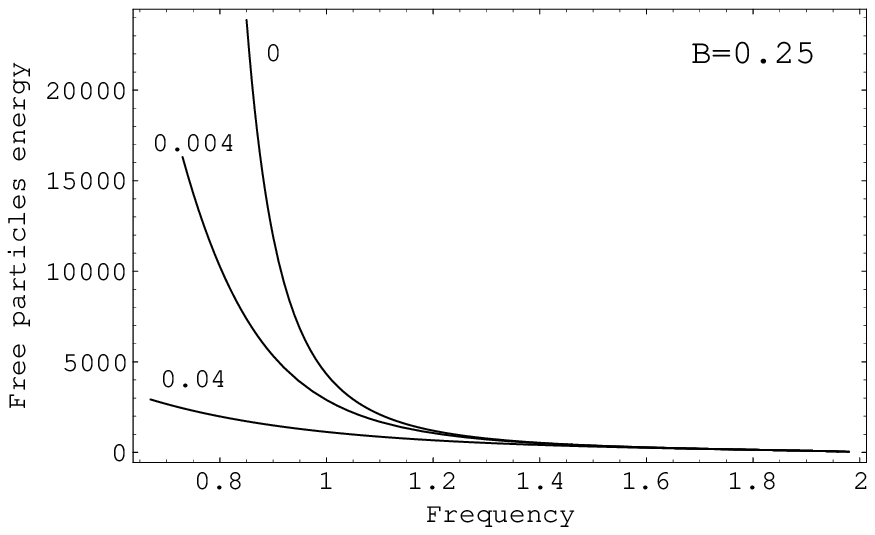}
\caption{The energy of the free particles with charge equal to
the soliton charge as a function of
$\sqrt{{\boldsymbol{\varepsilon}}^2}$ for three different values
of the coupling constant with $B=0.25$. The charge of the soliton
is half the value of the energy of the free particles.}
\label{figure3}
\end{figure}

\begin{figure}
\centering
\includegraphics{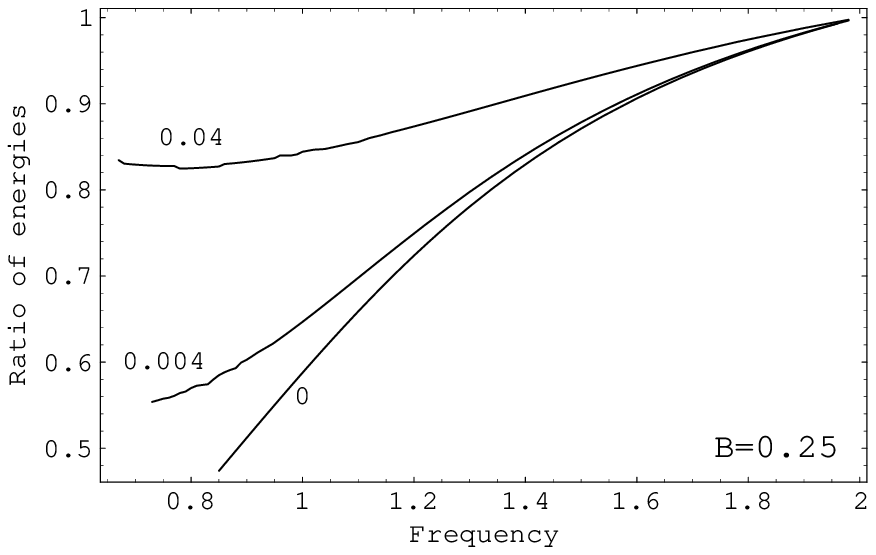}
\caption{The soliton energy per the energy of free particles with
equal charge as a function of
$\sqrt{{\boldsymbol{\varepsilon}}^2}$} \label{figure4}
\end{figure}

Figures \ref{figure1}, \ref{figure2}, \ref{figure3} and
\ref{figure4} show the behavior of the soliton as a function of
$\sqrt{{\boldsymbol{\varepsilon}}^2}$. The square of the coupling
constant takes the values 0, 0.004 and 0.04. The first is the case
of the global symmetry. We see that when
$\sqrt{{\boldsymbol{\varepsilon}}^2}$ approaches the critical
(minimum) value, $\sqrt{{\boldsymbol{\varepsilon}}_0^2}$, which
holds for the thin-wall approximation, the ``peak" of the soliton
approaches the $2/3B$ value according to eq. \ref{101}, and the
energy and the charge of the soliton increase. These all agree
with the idea that when the frequency approaches the critical,
minimum, value, the contribution of the spatial field derivatives
to the total soliton energy get smaller and we can talk for a
large soliton with large energy contributions from the time
dependence and the potential. When
$\sqrt{{\boldsymbol{\varepsilon}}^2}\rightarrow 2m$ the size,
energy and charge of the soliton decrease, tending to a flat field
configuration covering the whole space with a very small value
which corresponds to free particles. For the above reason the
ratio soliton energy per free particles energy approaches unity.
When the coupling constant differs from zero (i.e. when referring
to the case of local symmetry) the soliton energy and charge are
less than in the global case.

\begin{figure}
\centering
\includegraphics{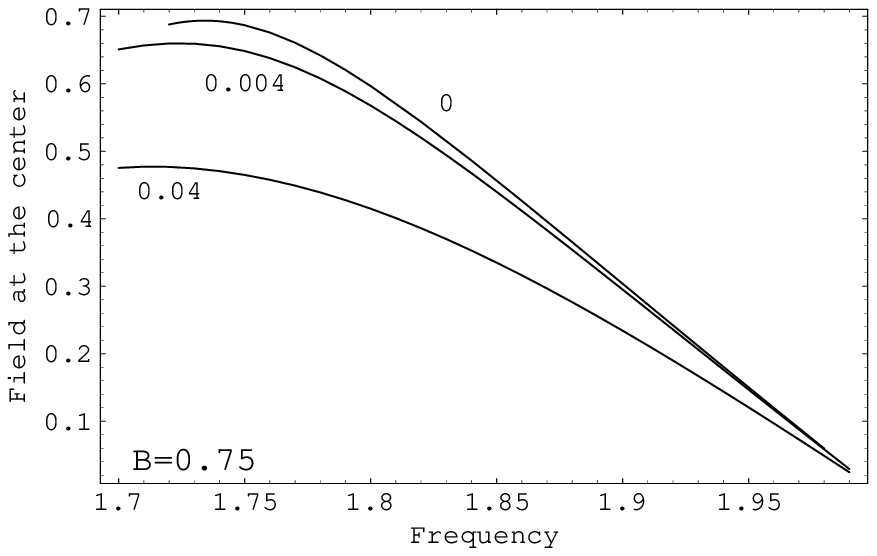}
\caption{The central value of the field as a function of
$\sqrt{{\boldsymbol{\varepsilon}}^2}$ for three different values
of the coupling constant with $B=0.75$.} \label{figure5}
\end{figure}

\begin{figure}
\centering
\includegraphics{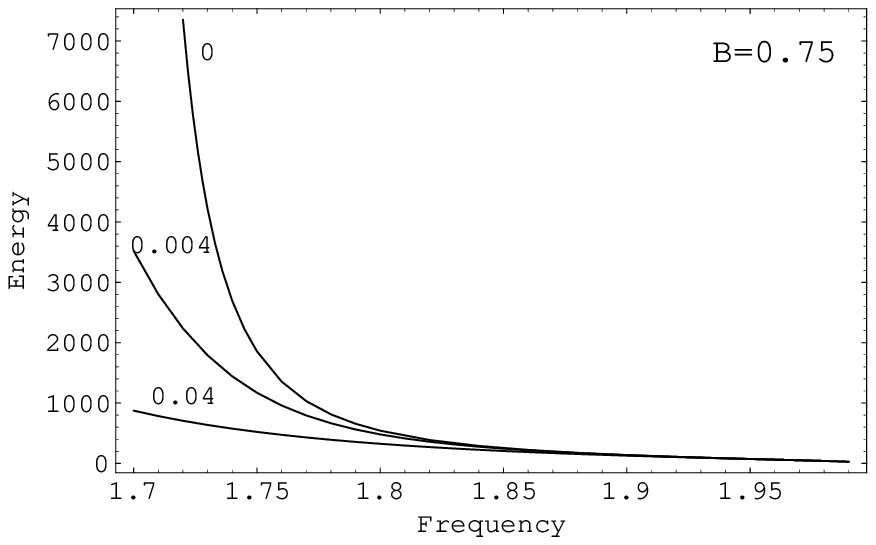}
\caption{The energy of the soliton as a function of
$\sqrt{{\boldsymbol{\varepsilon}}^2}$ for three different values
of the coupling constant with $B=0.75$.} \label{figure6}
\end{figure}

\begin{figure}
\centering
\includegraphics{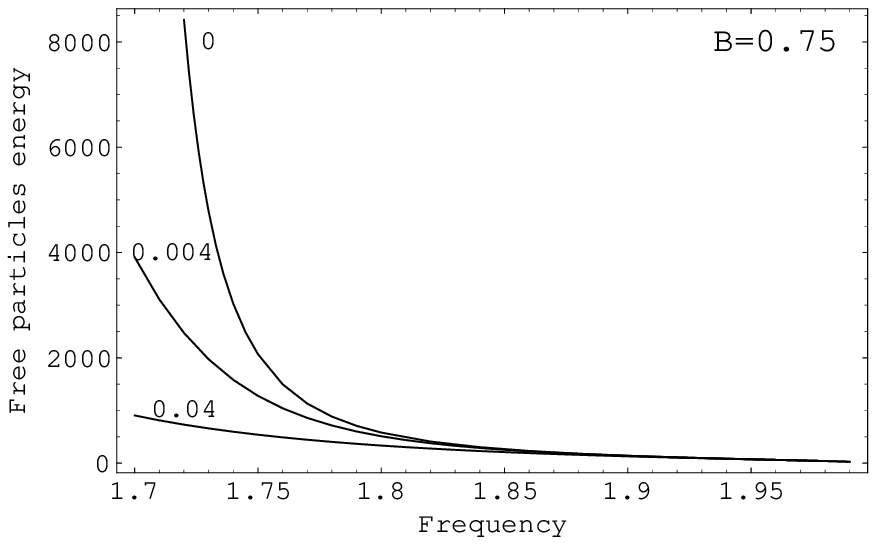}
\caption{The energy of the free particles with charge equal to
the soliton charge as a function of
$\sqrt{{\boldsymbol{\varepsilon}}^2}$ for three different values
of the coupling constant with $B=0.75$. The charge of the soliton
is half the value of the energy of the free particles.}
\label{figure7}
\end{figure}

\begin{figure}
\centering
\includegraphics{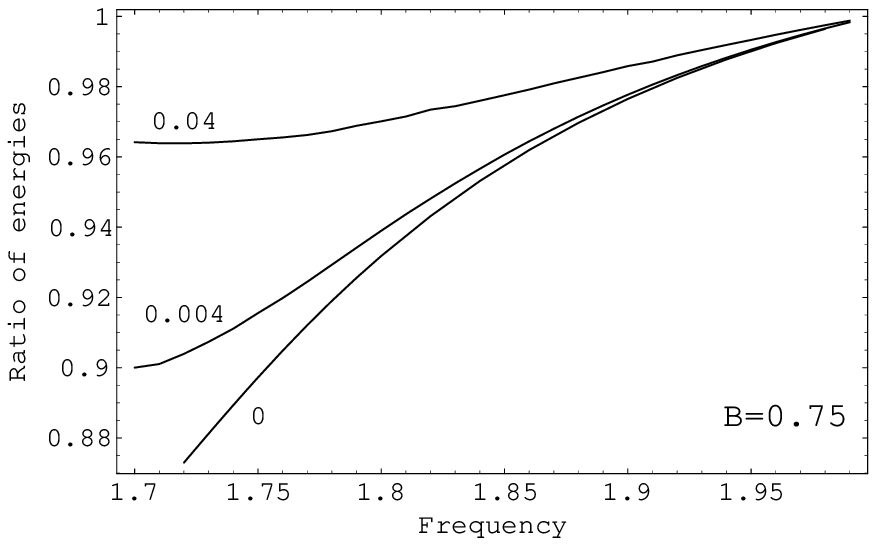}
\caption{The soliton energy per the energy of free particles with
equal charge as a function of
$\sqrt{{\boldsymbol{\varepsilon}}^2}$}.
\label{figure8}
\end{figure}

Figures \ref{figure5}-\ref{figure8} give the central value, the
energy of the soliton, the energy of free particles with the same
charge and the ratio of the two energies. The potential now is
less ``deep", as the new value of the parameter $B$ is 0.75. This
has some interesting consequences. The minimum value of the
frequency ($2\sqrt{1-2/9B}$) is now larger than in the case of
$B=0.25$. So the ratio of the soliton energy per the energy of the
free particles is in general larger than in the case of $B=0.25$,
because the minimum frequency is the lower limit of the above
ratio.

\begin{figure}
\centering
\includegraphics{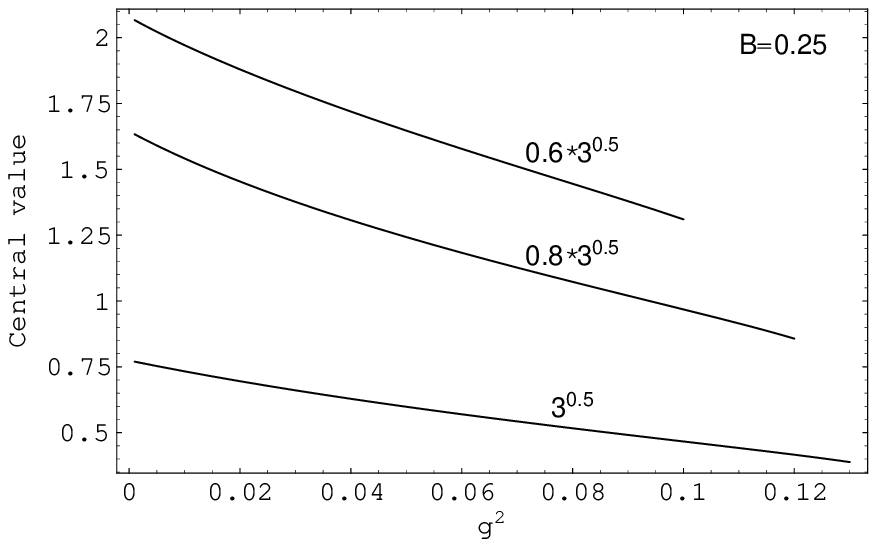}
\caption{The central value of the field as a function of the
coupling constant for three different values of the frequency with
$B=0.25$.} \label{figure9}
\end{figure}

\begin{figure}
\centering
\includegraphics{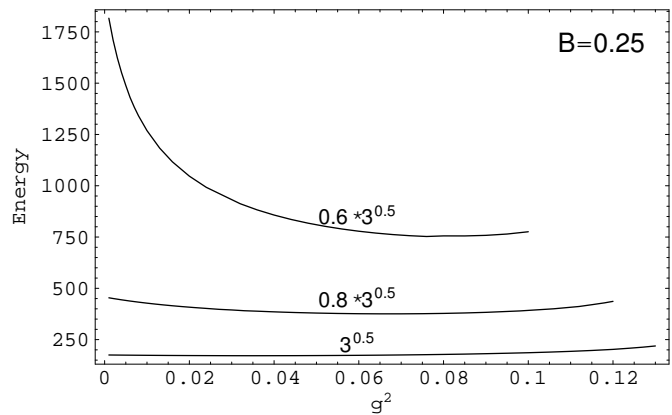}
\caption{The energy of the soliton as a function of coupling
constant for three different values of the frequency with
$B=0.25$.} \label{figure10}
\end{figure}

\begin{figure}
\centering
\includegraphics{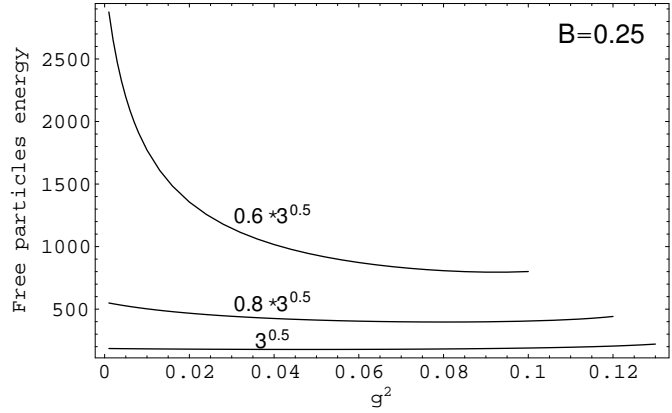}
\caption{The energy of the free particles with charge equal to
the soliton charge as a function of the coupling constant for
three different values of the frequency with $B=0.25$. The charge
of the soliton is half the value of the energy of the free
particles.} \label{figure11}
\end{figure}

\begin{figure}
\centering
\includegraphics{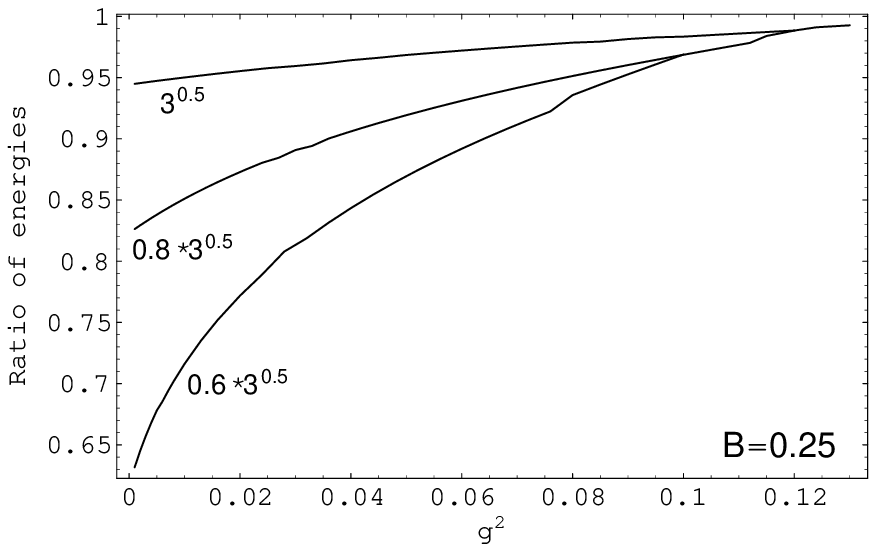}
\caption{The soliton energy per the energy of free particles with
equal charge as a function of the coupling constant with
$B=0.25$.} \label{figure12}
\end{figure}

In figures \ref{figure9}-\ref{figure12} we study the behavior of
the soliton when the coupling constant varies. When the coupling
constant increases the soliton value decreases. This is an
expected behavior due to the repulsion between the different
parts of the soliton and the main reason for the decrease of the
particle number (i.e. the soliton charge.) We will now explain
the behavior of the soliton energy as a function of the coupling
constant. Firstly the energy decreases with respect to the
coupling constant due to the decrease of the value of the scalar
field. Then, the energy seems to increase slightly due to the
increase of the ``electrostatic" energy contribution.

\section{Conclusions}

We found the energy and charge for a specific field configuration
corresponding to the q-ball ansatz (eq. \ref{6}) in a Lagrangian
of a spinor doublet of two scalar fields with a global $SU(2)$
symmetry. In the thin-wall approximation we found that, for
potentials raising near the origin as ${|\varphi|}^2$, then
slower than ${|\varphi|}^2$ and for large $|\varphi|$ faster than
${|\varphi|}^2$, q-ball-type solitons can be observed at or near
the minimum of the $\sqrt{U/{|\varphi|}^2}$ quantity. In this
case the soliton energy takes a simple form and is smaller than
the energy of the free particles with the same charge, provided
that the above quantity is less than the mass of the free
particles. In the thick-wall approximation we used the full
Euler-Lagrange equation for the field and after solving it
numerically we found that for a large region in the parameter
space the soliton is prevented from fission into free particles
thanks to the energy and charge conservation.

In the case of local $SU(2)$ symmetry we found that a localized
field configuration is possible as a solution of the
Euler-Lagrange equations. We found the energy of such a
distribution as a function of charge. We minimized the energy
with respect to the radius. Minimizing the energy with respect to
the field value we found that this value is less than the
corresponding to the global case due to the repulsion between the
different parts of the charged soliton. For small values of the
coupling constant and using numerical methods we found that
a)there are localized and, thus, non-topological-soliton-type
solutions to the equations of motion and b)their energy is less
than the energy of the free particles. The comparison of the
properties of the numerical solutions with different values of
the parameters (coupling constant, parameters of the potential
etc) helps us to fully understand the behavior of the soliton and
the differences between the local and the global case.

\vspace{1em}

\textbf{ACKNOWLEDGEMENTS}

\vspace{1em}

I wish to thank A. Kehagias, G. Leontaris and N. D. Tracas for
helpful discussions.

\end{document}